\documentclass[runningheads]{llncs}

\usepackage{caption}
\usepackage{subcaption}
\usepackage{todonotes}
\newcommand{\rr}[1]{\textit{#1}}
\usepackage{hyperref}
\usepackage{verbatim}
\usepackage{cite}

\begin{document}

\title{Code Review in the Classroom}

\author{Victor Rivera\inst{1} \and
Hamna Aslam\inst{2} 
\and
Alexandr Naumchev\inst{2}
\and
Daniel de Carvalho\inst{2}
\and
Mansur Khazeev\inst{2}
\and
Manuel Mazzara\inst{2}
}

\authorrunning{Victor Rivera et al.}
%
\institute{Australian National University, Canberra, Australia \\
\email{victor.rivera@anu.edu.au}\\
\and
Innopolis University, Innopolis, Russia\\
\email{\{h.aslam, a.naumchev, d.carvalho, m.khazeev, m.mazzara\}@innopolis.ru}}


\maketitle
\begin{abstract}
    This paper presents a case study to examine the affinity of the code review process among young developers in an academic setting. Code review is indispensable considering the positive outcomes it generates. However, it is not an individual activity and requires substantial interaction among stakeholders, deliverance, and acceptance of feedback, timely actions upon feedback as well as the ability to agree on a solution in the wake of diverse viewpoints. Young developers in a classroom setting provide a clear picture of the potential favourable and problematic areas of the code review process. Their feedback suggests that the process has been well received with some points to better the process. This paper can be used as guidelines to perform code reviews in the classroom.
\end{abstract}

\section{Introduction} \label{sec:introduction}
Professional Software Engineers and developers regularly implement code review, i.e., they review each other's code to identify errors, suggest improvements, and share knowledge in the team. Code reviews are useful and common in the industrial environment, e.g. they are part of the software development at Google and it is applied by more than 25K developers making more than 20K source code changes each workday \cite{google:2016}; however, they are not always applied as they should in academia, neither for what concerns research and development nor regarding pedagogical approaches. We consider code reviews a best practice to be applied in the academic environment.

Tools to support Code Reviews are abundant (see some of these in section \ref{sec:relatedwork}); in this paper, we are not introducing another tool of this kind, but we propose a process on how to perform a Code Review session in the university classroom, also in the presence of large groups. Our work follows the convergent practices across several code review processes and contexts identified in \cite{Rigby}: it is a lightweight and flexible process; reviews happen early (before a change is committed) and quickly; code sizes are small; few people are involved; and it is a group problem solving activity. We describe how this process can help in the development of the course and how it helps students to understand colleagues' mistakes and learn coding best practices. 

In particular, students can reflect on: 

\begin{itemize}
\item \textbf{Documentation}: Is the code properly documented and commented? (Possible answers: It needs more work / Somewhat / Yes it is great). Where should the documentation be better?
\item \textbf{Error handling}: Does the code handle errors properly? (Possible answers: It needs more work / Somewhat / Yes it is great). Where should the error-handling be better?
\item \textbf{Suggestions}: Provide two suggestions for the author on how to improve the code.
\end{itemize}

In this work, we focus on answering the following research questions:

\begin{itemize}
\item \textbf{RQ1}: \emph{What are the benefits and pitfalls for a course performing code reviews?}

Lecturers need to achieve specific learning outcomes for a course in a specific time frame. We want to explore how code review sessions can help lecturers in the process. We are also interested in exploring the consequences of adding code review sessions as part of the course. Finally, we want to determine whether there are more benefits than pitfalls: whether code reviews are worth in an academic setting.

\item \textbf{RQ2}: \emph{What are the benefits and pitfalls for students participating in code reviews?}

We want to determine whether code reviews help or hinder the learning process in students. This is important as it is directly related to students' motivation. The more motivated students are the more engaged they are in the activity \cite{Saeed:2012}, resulting in taking full advantage of the activity's benefits. There is also a need to understand what the pitfalls of the activity are so lecturers can focus on those elements. 

\item \textbf{RQ3}: \emph{What aspects of the artifacts under review do the students emphasise most?}

It is important to determine these aspects so lecturers, replicating the code review activity, can pay more attention to those elements. This makes the code review more focus, affecting (positively) the outcome of the activity.

\end{itemize}

The following high-level objectives motivate these research questions:
\begin{itemize}
    \item \emph{Keep the syllabus of the course realistic:} the course may not cover topics that should be covered.
    It may also loosely cover some of its topics, or cover some topics too extensively.
    We believe that conducting communication-intensive activities such as code review sessions may help uncover these problems. 
    \item \emph{Ensure good enough learning quality:} we may not deliver some topics of the course well enough.
    The course consists of both lectures and lab sessions.
    While the lectures may cover up-to-date theory, the labs may insufficiently or inadequately practice that theory.
    Some teaching assistants may practice the topics that they appreciate more.
    They also may have different backgrounds; this will likely affect their perception and delivery of the material.
    \item \emph{Ensure adequacy of the exercise itself:} the setting of the code reviews may require adaptation.
    It was the first time we applied code reviews for teaching first-year bachelor's.
    We designed the exercise based on how a code review would look like in a software company.
    The participants of code reviews in such companies understand very well the importance of code reviews.
    They also understand what aspects of the artifacts under review they should especially focus on.
    First-year bachelors most likely do not have that level of awareness.
    By studying their behaviour during code review sessions, we may learn how to adjust the setting of the exercise.
\end{itemize}

The rest of the paper is organised as follows. Section \ref{sec:relatedwork} gives a brief overview of the code review process in industry and academia, and shows different tools to perform reviews. Section \ref{sec:process} describes the process to perform code review in a classroom. The process describes both side of the activity: steps to be followed by educators and steps followed by students. Section \ref{sec:caseStudy} presents a case study where the process was applied to a large size course. Section \ref{sec:lessons} lists the lessons learnt after performing the code review. Finally, section \ref{sec:conclusions} is devoted to conclusions.

\section{Related Work}
\label{sec:relatedwork}
Code inspection consists of a manual review of the code by other developers than its authors to identify its defects. Already thirty-five years ago, it was recognised as a good engineering practice, which was successfully applied in several programming projects improving productivity and product quality \cite{fagan76}. At that time, it was done with in-person meetings: such code inspections can be called \emph{traditional} in contrast with practices that have been more recently introduced. In particular, \cite{BirdBachelli} defines \emph{Modern Code Reviews} as informal, tool-based and regular peer code reviews. Kemerer and Paulk \cite{Kemerer} analyse the impact of designing and performing code reviews in assignments of a Personal Software Process course written in C or C++: they empirically verify that allowing sufficient preparation time for reviews and inspections can produce better performance; this probably explains why doing such research in a teaching environment is easier than in an industrial one, which tends to try to decrease the short-term costs. Nevertheless, such research in an industrial environment exists: as an example, Sadowski \textit{et al.} \cite{CodeReview:Google} studies the practice of Modern code reviews at Google and its impact. Interestingly,  they do not only study the impact on the codes themselves but also on knowledge transfer due to code review; indeed, at Google, knowledge transfer is part of the educational motivation for code review. This educational capability of reviews was already emphasised in \cite{Johnson}.  Rigby and Bird \cite{Rigby} attempt to measure such spread knowledge across the development team through review. As recalled above, Modern Code Reviews in industry are tool-based. The pipeline followed by these tools (typically) starts by the owner of the code submitting the code to the tool. This allows reviewers to see the code and to propose changes (if needed); the tool enables both owner and reviewers to have discussions about the lines of the code. They can also propose changes on the code; finally, the code is patched or discarded. In an industrial setting, this is typically done before committing a change to the main repository. Some example of these tools are: CodeFlow \cite{codeflow:2018} used by Microsoft; Gerrit \cite{gerrit:2020} used by Google; ReviewBoard\cite{reviewboard:2014} used by LinkedIn and Mozilla; Phabricator\cite{phabricator:2020} used by Facebook and Quora. However, many aspects of the academic setting do not allow existing commercial code review processes to be used and present a software solution to code reviewing in academic settings so that students can get constructive comments about the code they turn in for assignments\cite{Tatarchenko12}. Code review in education has also been experimented in the context of secondary school \cite{Kubincova18}, where students perceived code review activities positively, and such activities helped in achieving good results in programming projects.

University lecturers who at the same time work in software production understand well the importance of code reviews, and realise the difficulties of teaching and implementing them in a class environment. Lecturers might not have a prior experience in presenting the topic, how to asses it, and even less how to coordinate the action over large groups. In this paper, we try to answer some of these questions. Some techniques applied in a university environment are described in \cite{Wind17}. The authors of the article discuss their approach and conclude that students actually learn from reviewing code and can assess quality, but also that generally, and understandably, do not like to grade the work of their peers. Hundhausen \textit{et. al.} presents a list of best practices for implementing code reviews in an academic setting\cite{CodeReview:2013}, the list is derived from empirical studies. Our work follows most of the practices: \textit{establishing ground rules} and \textit{modelling the activity} are part of the activities in phases (1) and (2) in figure \ref{fig:codeReview}. We did not, however, perform ``mock'' reviews beforehand; \textit{requiring both independent and team reviews} is part of step (2) in Phase (3) (figure \ref{fig:codeReview}); \textit{avoiding redundancy in reviewing} is achieved by the nature of our work as we suggest to code review in a more realistic scenario, students need to perform this activity before committing their code to a repository; we did not \textit{use trained moderators}. Moderators were trained, but they were students of the same course, we did not take into consideration their experience. 

\section{Code Review Process}
\label{sec:process}
Code review is a process in which a piece of code is manually inspected by developers other than the authors. The primary purpose of the code review is to improve the quality of software projects. Code reviews are frequent in the industry (e.g., \cite{CodeReview:Google}) where code is inspected before committing the changes in centralised repositories. Reviews ensure that quality code adheres to the company's best practices. The review process in the industry happens seamlessly as part of the software development; however, freshly graduated students are not prepared to perform the review (they instead learn the process on the fly). This section describes the process of performing a code review in an academic environment. 

\begin{figure*}[h]
\centering
\includegraphics[width=4.5in]{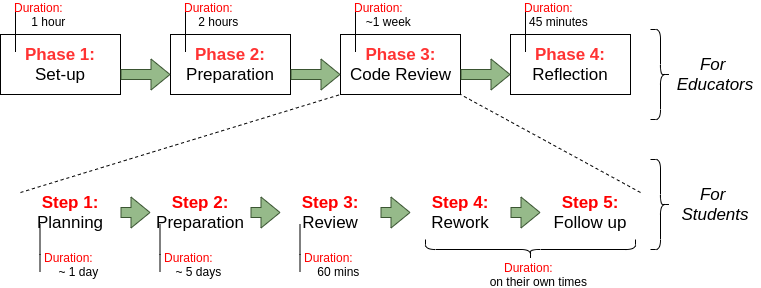}
\caption{Code Review process for the Classroom.}
\label{fig:codeReview}
\end{figure*}

The code review process defined in this paper assumes a project-based course that has at least six students, so at least one review session is performed. The process can handle large size courses; for instance, we applied it to a group of 129 students, a total of 22 reviews were performed (see section \ref{sec:caseStudy}). Figure \ref{fig:codeReview} depicts the whole process. The process has two main components (detailed in sections \ref{sec:codeReviewEducators} and \ref{sec:codeReviewStudents}). The upper part of the figure shows the process that educators performing the review need to follow, and the lower part the process of the actual code review. The process for educators starts with a set-up phase in which educators need to give directions and explanations about the review session to participants. In this phase, students form groups and roles are assigned. Phase 2, preparation, is about a series of meetings that educators perform with students in the same role (e.g., meeting with \rr{owner}s). In each meeting, roles are further explained to students, and some skills, e.g., facilitation for \rr{moderator}s, are exercised. Phase 3 is the actual Code Review; see below part of figure \ref{fig:codeReview}. It comprises five steps. It starts with planning the review. Here, roles for each code review are introduced, goals are defined, and the code for review is distributed. In step 2, preparation, \rr{reviewer}s review the unit of review off-line (as suggested in \cite{swinspection:1993}). There is an initial interaction between \rr{reviewer}s and the \rr{owner} to determine whether some of the review comments can be solved off-line. Step 3 is the actual review session. All participants, \rr{reviewer}s, \rr{owner}s, and \rr{moderator}s meet to discuss only those review comments that could not be taken off-line. Finally, Rework, and Follow-up steps take place. During the Rework step, the \rr{owner} implements the review comments that were not solved. During the Follow-up step, \rr{owner}s notify everyone about the changes. Once the code review is done, a Reflection phase is performed (Phase 4). In this phase, educators and participants reflect on the whole activity. 

\subsection{Components of the Code Review}
\subsubsection{Artifacts}
An artifact is one of many tangible products produced during the development of software. It can be any documentation of the software, for instance, UML diagrams, Internal/External documentation. It could also be an agile artifact from an Agile methodology, such as User Stories, acceptance criteria. Or it could be a piece of code such as a class (or a small number of closely related classes), a setup script, test suites. The artifact in the Code Review process plays a crucial role as it is the unit of review. The owner of the Code Review selects it, and it is the unit that reviewers check. The artifact of Code Reviews in companies is typically a piece of code that needs to be committed.

The artifact should not be extensive, as this would impact (negatively) the effectiveness of the review. It should be around 10 pages of text, for artifacts of software documentation, or around 250 LOC for artifacts of software code. The process described in the present paper assumes an artifact of software code.

\subsubsection{Roles}
There are three leading roles in a Code Review (roles are mutually exclusive, e.g., an \rr{owner} cannot be at the same time a \rr{reviewer} or a \rr{moderator}). 

\paragraph{\rr{Owner}s} They are the owners of the artifact to be reviewed. Their role is to provide the artifact of the review, along with an explanation.  \rr{Owner}s receive comments from \rr{reviewer}s about the artifact. Each comment needs to be addressed. \rr{Owner}s can defend their code and reject the comment, or accept the comment and work on it.

\paragraph{\rr{Reviewer}s} The set of people reviewing the artifact. Their role is to read carefully the artifact provided by the \rr{owner}s and make comments on it based on the review scope (see below). If \rr{owner}s require further explanation about the comments, \rr{reviewer}s are required to provide it. 

\paragraph{\rr{Moderator}s} They organise the review meeting. \rr{Moderator}s need to keep \rr{reviewer}s focused on the process (as \rr{reviewer}s might not be doing their job, this is common in the industry) and they also need to keep the discussion on track. After the review is done, \rr{moderator}s need to ensure that the follow-up step happens. This is important to keep the effectiveness of the process. During the review, \rr{moderator}s need to have good skills of facilitation: not just moderating the review but also correcting inappropriate behaviour; they should be impartial, without any judgement. At the end of the review, \rr{moderator}s prepare a report.

\subsubsection{Review Scope}
The review of the artifact should have a specific scope so that \rr{reviewer}s can focus on the specificity of the code. Bertrand Meyer\cite{Meyer:CodeReview:08} defines the following review scope sections, from high-level to more implementation-oriented:

\begin{enumerate}
\item \textit{Design decisions}: participants will discuss the high-level design of the artifact. For instance, is a certain class really justified or should have its functionalities merged with another class?;
\item \textit{API design}: participants will discuss the usability of software elements by others, and in particular to reuse. For instance, do any derived classes have common members that should be in the base class?;
\item \textit{Architecture}: participants will discuss the architecture where the artifact sits. For instance, should a class really inherit from another or instead be a client?;
\item \textit{Implementation techniques}: participants will discuss choices of data structures and algorithms. For instance, can better data structures or more efficient algorithms be used?;
\item \textit{Exceptions handling -- Contracts}: participants will discuss how the artifact handles unexpected behaviours or about the use of code contracts. For instance, have all array (or other collection) indexes been prevented from going out-of-bounds? are all object references initialised before use? If not, how this is being handled?;
\item \textit{Programming style, names}: participants will discuss the conventions of the code style and conformance to standards and specifications. For instance, are descriptive variable and constant names used in accord with naming conventions? are there literal constants that should be named constants?;
\item \textit{Comments and documentation}: is documentation complete, including Design By Contract or Error checking specs as appropriate? are error messages comprehensive and provide guidance as to how to correct the problem?.
\end{enumerate}

\rr{Reviewer}s' comments will belong to one of these seven sections. Each comment will be tagged by \rr{owner}s with one of the following categories:

\begin{description}
\item[\texttt{[CLOSED]}:] there was a clarification about the comment, and the owner (with the \textit{reviewer}'s consent) decides to close the comment. This comment does not need any follow up;
\item[\texttt{[TO BE IMPLEMENTED]}:] the \textit{owner} agrees with the comment and decides to take action on it. These comments need to be implemented only after the review meeting;
\item[\texttt{[TO BE DISCUSSED]}:] after a short discussion about the comment, \textit{owner} and \textit{reviewer} reach an impasse. Comments with this tag are the ones being discussed during the actual code review meeting. 
\end{description}

\subsection{Code Review (for educators)}
\label{sec:codeReviewEducators}
This section describes the process that educators need to follow to apply a Code Review Session. The process contains four phases: Set-up, Preparation, Code Review and Reflection (upper part of figure \ref{fig:codeReview}). Phases 1 and 2 could be shortened after the first time the process is done. In a project-based course, the code review can be done several times. After the first time around, there is no need to spend time setting up the group and in preparation. Although it is advised to revisit those phases.

\subsubsection{Phase 1: Set-up}
In Phase 1, educators need to talk to all students involved in the process. An important part of the setting up is to motivate the process, so students engage with the activity. It is important as a great part of the success of this activity is the level of engagement of its participants. This phase also comprises the explanation of the process. Starting with the different roles (i.e. \rr{owner}s, \rr{moderator}s and \rr{reviewer}s), the review unit and the review scope. Finally, the process to follow (as explained in section \ref{sec:codeReviewStudents}). 

After the corresponding explanation, review groups are formed. Groups will have between five to seven participants: the \rr{owner} of the artifact, the \rr{moderator} of the process, and between three to five \rr{reviewer}s. It is essential to have at least three \rr{reviewer}s, so the feedback about the artifact is meaningful, and less than five \rr{reviewer}s so the review is kept within limits (the length of the sessions increases per participant). In case the project being developed in the course is in groups, it is advised that the groups of the code review are not the same. This is to avoid biased comments.

Phase 1 might take around one hour. 

\subsubsection{Phase 2: Preparation}
Three meetings take place in Phase 2: one meeting per role. Participants in the specific role must attend their role meeting. Although they can attend all three meetings. Each meeting starts by explaining the specific role. Part of this explanation was already given (Phase 1); the difference is that the audience is more focused. Several examples of code reviews are shown so they get an initial idea of how they should perform theirs. At the end of the meeting, there is a QA session. Each meeting has some specific discussions:   

\begin{enumerate}
    \item meeting with \rr{owner}s: educators should mention how feedback should be received. Some of the students might not be comfortable receiving negative feedback. It is important to remind them that this activity is for their benefit and that no one is trying to do harm. Educators should also mention that \rr{owner}s need to be responsive (answering to comments on a regular basis) as this affects the effectiveness of the review. Finally, \rr{owner}s should be open-minded about the comments of their peers.
    
    \item meeting with \rr{reviewer}s: educators should mention how feedback should be given. This might be the first time for some students to give feedback. Feedback should 
    \begin{itemize}
        \item be descriptive and not evaluative. The more specific, the better. If a reviewer finds a problem in the code, they should say where exactly and how it can be reproduced. They can also suggest how to solve the problem;
        \item take into consideration the needs of the artifact (do not show off how much better/smarter you are);
        \item be balanced. It is important to acknowledge \rr{owner}'s effort as well;
        \item be sensitive to \rr{owner}s: instead of ``you didn't initialise this variable'' you could write ``I didn't see where this variable was initialised''.
    \end{itemize}
    \item meeting with \rr{moderator}s: educators should mention the relevance of moderating a session. They should also explain some of the techniques for facilitating a session and resolving conflicts. It is also important to tell them that a \rr{moderator} should be impartial and should not impose their judgement. 
\end{enumerate}

Each meeting in phase 2 takes around 40 minutes.

\subsubsection{Phase 3: Code Review}
Steps of phase 3 are explained in detail in section \ref{sec:codeReviewStudents}.

\subsubsection{Phase 4: Reflection}
Once the review meeting (Step 3 in phase 3) takes place, there is a Reflection (Phase 4) session. The reflection session is performed to help students organise their thoughts around the activity. All participants (in all roles) are asked to reflect upon the review process activity. This session follows the Focused Conversational Model\cite{orid}. The model enables a conversation to flow surface to depth. It follows the Objective-Reflective-Interpretative-Decisional (ORID) method for asking questions. This model is being practised in various domains to examine an individual's perception of the process they are involved in and their corresponding affordances, primarily investigations on design interactions \cite{aslam2020affordance}.

\paragraph{Objective questions: }
The purpose of these questions is for participants to think about the facts of the activity. There is no interpretation or feeling involved at this stage. Participants should recall all steps of the activity, and what was observed there, this gives the context. Example of \textit{Objective} questions are

\begin{itemize}
\item Could anyone recall the steps of the review process? What was the process of each step?
\item What was the role of a \rr{moderator}?
\item Could you please describe what was this code review session about?
\item How many comments were discussed during the review?
\end{itemize}

\paragraph{Reflective questions: }
The purpose of these questions is to evoke immediate personal reactions, emotions, feelings, or associations. Notice that it is expected to have similar answers for \textit{Objective} questions, but answers for \textit{Reflective} questions are personal. Example of \textit{Reflective} questions are

\begin{itemize}
\item How did you, as \rr{owner}, feel when you receive specific feedback?
\item Can you associate those emotions with something that has happened in your life?
\item What do you feel when the \rr{owner} marked a comment as \texttt{[CLOSED]}, but you think it was not?
\item How did you, as \rr{moderator} feel when a situation was out of control?  
\end{itemize}

\paragraph{Interpretative questions: }
The purpose of these questions is to draw out meaning, significance, or implications about those reactions. Example of \textit{Interpretative} questions are

\begin{itemize}
\item What new insight did you, as \rr{moderator}, get from the activity?
\item What ways of feedback work better for you?
\end{itemize}

\paragraph{Decisional questions: }
The purpose of these questions is to bring the conversation to a close, eliciting resolution and enabling participants to make a decision. Example of \textit{Decisional} questions are

\begin{itemize}
\item What are two or three key points that you take away from this activity?
\item What have you learned from this activity?
\item Do you think you will be using Code Review? Regularly?
\item Do you have any suggestions for improving the activity?
\end{itemize}

Phase 4 might take around 30 minutes.

\subsection{Code Review (for students)}
\label{sec:codeReviewStudents}
This section describes the process that participants need to follow, the actual Code Review process. The process contains five steps: Planning, Preparation, Review, Rework and Follow-up (as depicted in the bottom of figure \ref{fig:codeReview}).

\subsubsection{Step 1: Planning}
The \rr{owner} decides on an artifact and freezes it in the repository. The artifact cannot be modified during the review process (this might affect \rr{reviewer}s' comments). The artifact needs to be ready for comments at least five days before the actual meeting.

\begin{figure}[h]
\centering
\includegraphics[width=3in]{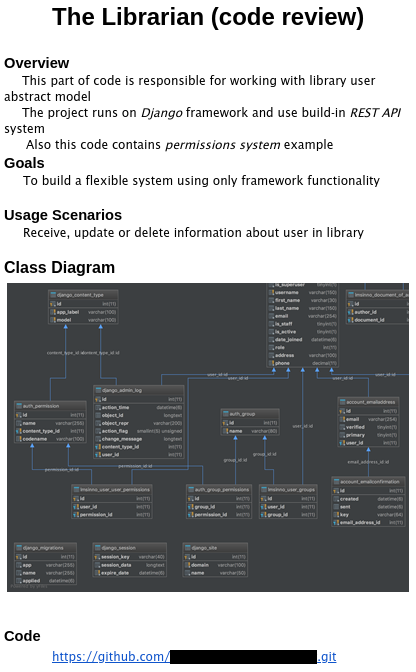}
\caption{Code review template. Google document excerpt}
\label{fig:google}
\end{figure}

The \rr{moderator} determines the objective of the review session and shares it with participants. Objectives of the review in a classroom need to be aligned with the objectives of the course. \rr{Moderator}s are in charge of determining the objectives, but there is a strong influence from the educator of the course. The \rr{moderator} also creates a shared Google document in which \rr{owner}s write a short overview of the artifact, along with some useful information for its understanding, e.g., Usage Scenarios, UML diagrams where the artifact sits. The description should not be a full description of the project, rather a brief overview of the artifact. The Google document should also have a link to the artifact's repository. The \rr{moderator} grants write permissions to all \rr{reviewer}s and the educator of the course and distribute the link among them.  Figure \ref{fig:google} shows an example of the shared Google document.

\subsubsection{Step 2: Preparation}
Once the shared document is distributed, \rr{reviewer}s should read the artifact and start commenting on the Google doc. The main idea is to have an initial interaction with the \rr{owner}. This activity lasts for five days, one day before the actual review meeting is held. \rr{Reviewer}s comment on the artifact based on the review scope, i.e., design decisions, API design, architecture,  implementation techniques, exception handling, programming style, and documentation.  \rr{Reviewer}s are not supposed to run the code. Their comments should be solely based on the artifact in isolation. 

\begin{figure*}
     \centering
     \begin{subfigure}[b]{0.6\textwidth}
         \centering
         \includegraphics[width=\textwidth]{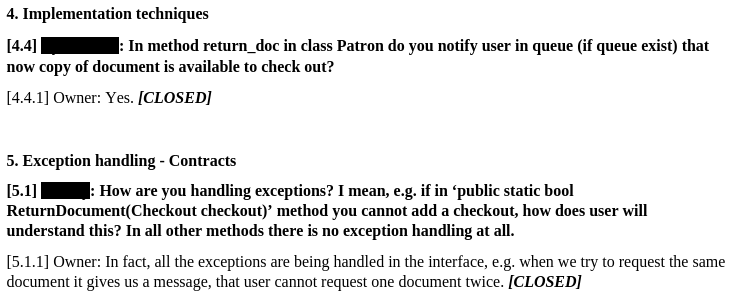}
         \caption{\textit{\textbf{[Closed]}} tag}
         \label{fig:closedtag}
     \end{subfigure}
     
     \begin{subfigure}[b]{0.6\textwidth}
         \centering
         \includegraphics[width=\textwidth]{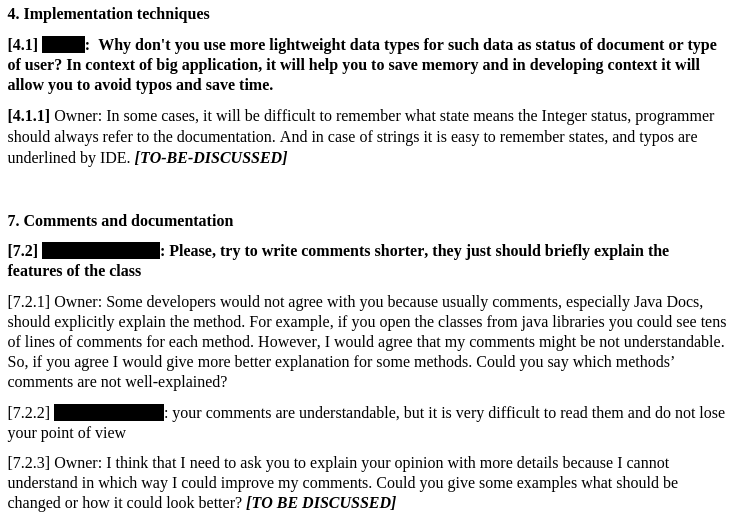}
         \caption{\textit{\textbf{[TO BE DISCUSSED]}} tag}
         \label{fig:tobediscussedtag}
     \end{subfigure}
     
     \begin{subfigure}[b]{0.6\textwidth}
         \centering
         \includegraphics[width=\textwidth]{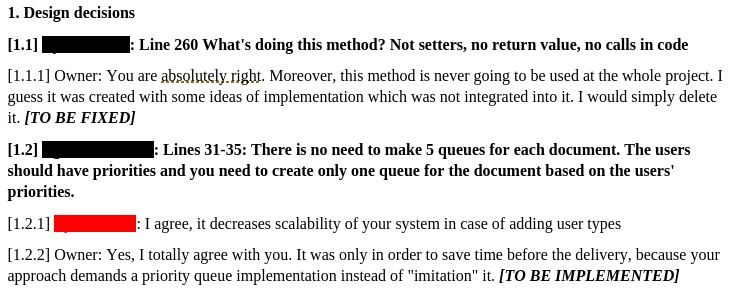}
         \caption{\textit{\textbf{[TO BE IMPLEMENTED]}} tag}
         \label{fig:tobeimplementedtag}
     \end{subfigure}
        \caption{Excerpt of Code Review tags (and variants)}
        \label{fig:tags}
\end{figure*}

During these five days, \rr{reviewer}s will comment on the artifact, and the \textit{owner} will reply to their comments. The purpose is to reduce the number of topics to be discussed during the actual meeting. Some of the comments can be cosmetic; some others need no discussion as an agreement can be found. For this, the \rr{owner} will try to clarify the comments and will tag each comment in the Google doc with one of the following categories \texttt{[CLOSED]}, \texttt{[TO BE IMPLEMENTED]}, \texttt{[TO BE DISCUSSED]}. An example of such comments with different tags is shown in Figure \ref{fig:tags}.

\subsubsection{Step 3: Review}
The actual review meeting is held. \rr{Reviewer}s, \rr{owner} and \rr{moderator}  sit together for around 60 minutes (it could be as long as 90 minutes). The meeting starts with an introduction of all participants (so everyone knows their role), then a statement of objectives (so everyone is on the same page).  During the meeting, \rr{owner} and \rr{reviewer}s discuss those comments in the Google doc that require attention, those tagged as \texttt{[TO BE DISCUSSED]}. The \rr{moderator} facilitates the meeting, making sure all comments are addressed, all \textit{reviewer}s participate, and no one monopolises the review. 

\subsubsection{Step 4: Rework}
This step of the process is offline. The \rr{owner} of the artifact should investigate the issues in the Google document tagged as \texttt{[TO BE IMPLEMENTED]} and implement them, or at least report them as issues in a tracking system. This is important not just because the \rr{owner} fully takes advantage of the Code Review, but also because \rr{reviewer}s can see the progress on the comments they had spent time on.

\subsubsection{Step 5: Follow up}
The final step is the Follow Up. The \rr{owner} should report on the Google document the results of the Code Review. Participants can access the doc to confirm that fixes have been implemented. The \rr{moderator} can collect data such as the number of defects,
number of participants, number of fixes, and saved as issues, and total time spend reviewing. This can be used to improve any upcoming review process.

\section{Case Study}
\label{sec:caseStudy}
This section reports on the Code Review process performed in a large size course of a Computer Science program. 

\subsection{Course Structure}
The \textit{Introduction to Programming II} course at Innopolis University is a 6 ETCS course delivered to freshmen in the second semester over 15 weeks with 2 academic hours of frontal lectures and 2 hours of laboratories every week. This course is a continuation from \textit{Introduction to Programming I}, a course that focuses in Object-Oriented Programming and the notion of Software Contract using the metaphor of business contract\cite{teachingDbc:18}. Design by Contract\cite{Meyer:1997} is an approach to achieve the so-called Correctness by construction\cite{Chapman:2006}. Correctness by Construction makes use of foundations of logic, concepts that are taught by a \textit{Discrete Math} course. \textit{Introduction to Programming II} is a project-based course. After successfully taking this course, students will master the fundamental data structures and algorithms, modular programming, exception handling, and programming language mechanisms, as well as the fundamental rules of producing high-quality software. The teaching team is composed of a Principal Instructor (PI) in charge of delivering lessons and Teaching Assistants (TA) in charge of delivering lab sessions. Around 90\% of the first-year students are between 17 and 19 years old (the rest are no older than 31), due to the specific structure of the Russian scholastic itinerary and around 80\% of the students are Russians.

Students of \textit{Introduction to Programming II} needed to implement a project where they had to apply all concepts learnt. The project was the classical Library Management System (LMS). LMSs are used in libraries to track the different items of the library, e.g., books, magazines, audio/video materials. The system also keeps track of people allowed to check out those items (patrons) and people in charge of the management of the library (librarians). The project was divided into four deliveries, and students were free to use any programming language and paradigm, and any Framework, the only restriction was that they had to host the source code in any subversion repository (so the PI could check the progress of each member of the team). The course size was 193 students, and they had to work in groups of around six members (they were free to select their teams). Along with the third delivery, students had perform a Code Review activity.

\subsection{Code Review Phases}
\paragraph{Phase 1: Set-up} Students were informed about the Code Review process. The PI of the course explained to them the different elements of the review, the artifacts, the roles, and the review scope, along with the tags for comments. All the process was explained. They were instructed to form teams of five to seven participants and to select the role of each participant (i.e. \rr{owner}, \rr{moderator} and \rr{reviewer}s). They were advised to form different groups for the review than the groups of the project. This phase was performed in 1.5 hours.

\paragraph{Phase 2: Preparation} Once all groups were formed and roles were decided, three different meetings were scheduled. These meetings happened outside the lecture time. All meetings were open to all participants. During these meetings, the PI explained in more detail the process of the review and each role. Examples of code reviews were shown so participants could get an idea of the process. Each meeting lasted for about 30 minutes. The longest one was with \rr{moderator}s as this role is less natural for students: they had not moderated a meeting before. 

\paragraph{Phase 3: Code Review} 
As the size of the course was large, the actual Code Review meetings were done in parallel with the help of the Teaching Assistants (TA) of the course. They were instructed to spend the first ten minutes of the session, reminding students about the activity. This includes explaining the logistics, how the meeting will be conducted, and the duration (60 minutes) as well as the different roles of the review and a reminder to moderators that they should be facilitating the meeting. 

Students were split among their review teams and placed them around the classroom. The classroom should be big enough to allow at most three different teams to discuss without affecting each team. Then the review starts: the \rr{moderator} starts facilitating. During the review, TAs cannot influence nor participate in the discussions. Although, TAs should be alert in case there is misconduct. In such a case, TAs can intervene just to guide them to reach consensus rather than impose an opinion. TAs notify students when there are ten minutes left to finalise the review.  More details about the review are shown in section \ref{subsec:students} below.

\paragraph{Phase 4: Reflection}
As the Code review was done in parallel with the help of TAs, TAs were instructed on how to conduct the reflection session. After the review meeting was over, TAs gathered all students and reminded \rr{owner}s to take care of the Follow-up part of the exercise. As well as to update the document once they take care of issues. The reflection session is for students to think and reflect upon the activity that they were exposed. TAs conduct this activity using the ORID method. TAs were also instructed to make students participate in the activity, e.g., by asking direct questions. This activity lasted for 30 minutes.

\subsection{Code Review steps}
\label{subsec:students}

\paragraph{Step 1: Planning} \rr{Owner}s were instructed to freeze the artifacts and to produce a description of the artifact. \rr{Moderator}s were instructed to determine a set of objectives for the review. The PI influenced this process as the objectives should be aligned with those of the course. The objectives of the review were set to answer the following questions.

\begin{itemize}
    \item Does the artifact make use of the appropriated data structures? Is the code scalable?;
    \item What is the level of quality of the artifact presented? Is the owner using Design by Contract mechanisms? are unexpected behaviours being handled (Exceptions)?;
    \item Does the artifact follow a defined architecture?;
    \item Is the artifact maintainable?.
\end{itemize}

\rr{Moderator}s were also instructed to create a Google document and share it (with write permissions) to all \rr{reviewer}s and the PI (for monitoring). The document should contain \rr{owner}'s description of the artifact and the link to the artifact.

\paragraph{Step 2: Preparation}
There are five days in which \rr{reviewer}s review the artifact and make comments on the Google document. \rr{Owner}s need to continuously answer to those comments, tagging them as either \texttt{[CLOSED]}, \texttt{[TO BE IMPLEMENTED]} or \texttt{[TO BE DISCUSSED]}. During this time, \rr{moderator}s need to make sure that the process is in place and that all participants are working on it. As the PI has permission to the shared documents, the PI should continuously monitor the progress of this step. PI intervenes only if no one is working on the process. It is advised for the PI to give feedback on how to write comments. This is to reduce ambiguity. 

\paragraph{Step 3: Review}
The actual meeting happened. As explained before, this was a great size course, so the reviews were monitored by the PI and TAs. They were supposed to intervene only in case something went out of control. Nothing damaging happened during the sessions. Some groups had more discussions than others, but overall respect governed the process. All shared documents are available in \cite{Results:CodeReview} (\textit{Code Reviews} folder). Names were changed to protect the identity of students, and links to their code were removed. The reader can used these review documents to show students before performing a Code Review session in their courses.

\paragraph{Step 4 and 5: Rework and Follow up}
\rr{Moderator}s were instructed to monitor the progress of the work after the review meeting. They had to check whether \rr{owner}s were updating the shared document. If \rr{owner}s did not update the document, \rr{moderator}s could send a reminder to them. The PI checked all shared documents at the end of the course, and all comments were tagged as \texttt{[CLOSED]}; some of those comments were reported as tickets to be taken care of in the future. 

\rr{Moderator}s also collected data about the number of bugs found offline and during the review, and how many of them resulted in false bugs. 

\subsection{Data/Observations}
There were a total of 129 students who participated in the activity.
There were 22 groups for code review, hence 22 \rr{owner}s and 22 \rr{moderator}s.
19 groups had 4 \rr{reviewer}s and the rest had 3 \rr{reviewer}s.
The artifacts contained, on average, 200 Lines of Code.
After the review process, \rr{moderator}s reported that \rr{owner}s found a total of 122 bugs during off-line (before the review meeting) and 19 bugs during the review.
There were a total of 656 interactions on the Google documents between \rr{owner}s and \rr{reviewer}s.
All Google docs can be accessed in \cite{Results:CodeReview} (\textit{Code Reviews} folder).
Table \ref{table:commentscount} shows the different categories on the Google docs and the number of comments tagged as \texttt{[CLOSED]}, \texttt{[TO BE IMPLEMENTED]} and \texttt{[TO BE DISCUSSED]}.
The bottom row and rightmost column show the totals.
Around 41\% of the comments were tagged as \texttt{[CLOSED]}: \rr{owner}s explained the issues offline and decided to take no action; around 52\% of the comments were tagged as \texttt{[TO BE IMPLEMENTED]}: \rr{owner}s decided to implement \rr{reviewer}s' comments after the review process is finished; and only 21 comments (around 7\%) \texttt{[TO BE DISCUSSED]}.
The review meeting focuses only on these comments, so the meeting is very concrete, making it more effective.
We have found out that there are fewer comments about the high-level aspect of the artifact (categories (1) to (3)) than about the implementation-oriented aspects (categories (4) to (7)).
The reason is that it is more difficult to focus on high-level aspects by looking only at a piece of code.
Around 30\% of the comments were about the high-level aspect of the artifact and 70\% about the implementation-oriented aspects.

\begin{table*}[]
\begin{tabular}{lcccc}
\textbf{Category}                        & \multicolumn{1}{l}{\textbf{CLOSED}} & \multicolumn{1}{l}{\textbf{TO BE IMPLEMENTED}} & \multicolumn{1}{l}{\textbf{TO BE DISCUSSED}} & \multicolumn{1}{l}{} \\ \cline{1-4}
\multicolumn{1}{|l|}{(1) Design Decisions}   & 19                                  & 20                                             & \multicolumn{1}{c|}{5}                       & 44                   \\
\multicolumn{1}{|l|}{(2) API Design}         & 10                                  & 7                                              & \multicolumn{1}{c|}{2}                       & 19                   \\
\multicolumn{1}{|l|}{(3) Architecture}       & 10                                  & 12                                             & \multicolumn{1}{c|}{3}                       & 25                   \\
\multicolumn{1}{|l|}{(4) Impl. Techniques}   & 24                                  & 27                                             & \multicolumn{1}{c|}{2}                       & 53                   \\
\multicolumn{1}{|l|}{(5) Exception Handling} & 16                                  & 23                                             & \multicolumn{1}{c|}{3}                       & 42                   \\
\multicolumn{1}{|l|}{(6) Programming Style}  & 22                                  & 47                                             & \multicolumn{1}{c|}{3}                       & 72                   \\
\multicolumn{1}{|l|}{(7) Comments and Doc.}  & 21                                  & 22                                             & \multicolumn{1}{c|}{3}                       & 46                   \\ \cline{1-4}
                                         & 122                                 & 158                                            & 21                                           &                     
\end{tabular}
\caption{Code review categories and number of comments per tag.}
\label{table:commentscount}
\end{table*}

Table \ref{table:commentscount} also shows that the most common comments were on \texttt{Programming Style}.
This was expected as \rr{reviewer}s only have access to the artifact and could not run the code.
Most of the comments, around 65\%, are to be implemented.
This is followed by comments on \texttt{Implementation techniques} and \texttt{Comments and Documentation}.
The behaviour was expected and went in the direction of the Code Review practices at Google\cite{CodeReview:Google}.
Categories with the least comments were \texttt{API Design} and \texttt{Architecture}. It is challenging to comment on these two categories having only the artifact. Typically, artifacts are not large (otherwise, the effectiveness of the review is affected\cite{ibm:2011}), so \rr{reviewer}s cannot see the big picture. Although, in these two categories, around 42\% of the comments were tagged to be implemented. Meaning, even though these categories are difficult to comment on, those few comments were necessary. 

After the code review activity, \rr{reviewer}s were asked to fill up a questionnaire about the reviewed code.
98 of the \rr{reviewer}s participated in the survey.
The survey contained four \textit{Likert} questions (linear scale from 1 to 5 -- from \texttt{Strongly Disagree} to \texttt{Strongly Agree}):

\begin{enumerate}
    \item Is the reviewed code modular?
    \item Does the reviewed code implement a proper logging mechanism?
    \item Is the reviewed code understandable?
    \item Does the reviewed code handle exceptions?
\end{enumerate}

\begin{figure*}
     \centering
     \begin{subfigure}[b]{0.4\textwidth}
         \centering
         \includegraphics[width=\textwidth]{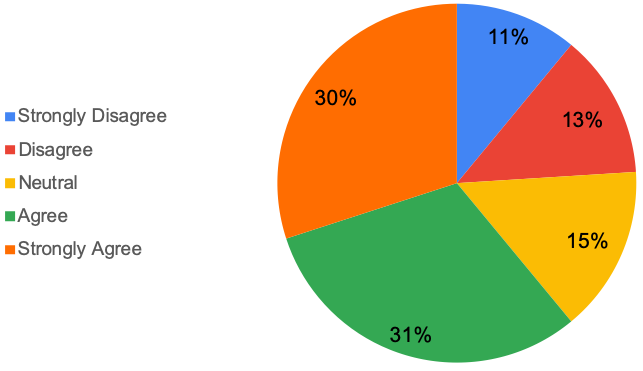}
         \caption{Is the reviewed code modular?}
         \label{fig:q1}
     \end{subfigure}
     \hfill
     \begin{subfigure}[b]{0.4\textwidth}
         \centering
         \includegraphics[width=\textwidth]{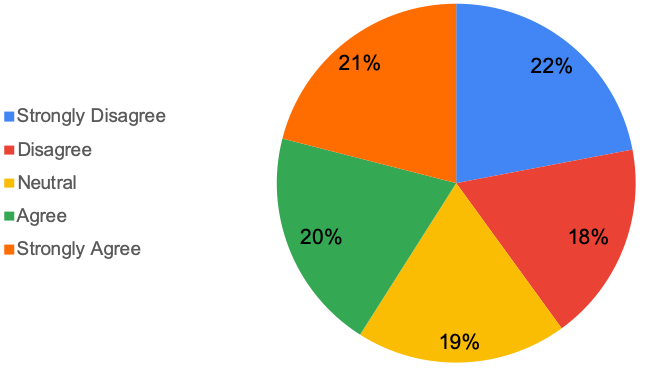}
         \caption{Does the reviewed code implement a proper logging mechanism?}
         \label{fig:q2}
     \end{subfigure}
     
     \begin{subfigure}[b]{0.4\textwidth}
         \centering
         \includegraphics[width=\textwidth]{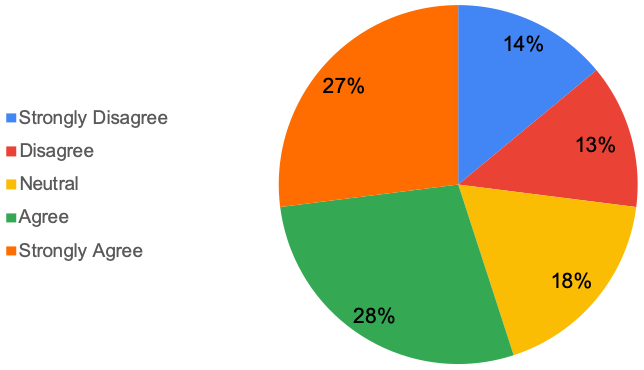}
         \caption{Is the reviewed code understandable?}
         \label{fig:q3}
     \end{subfigure}
     \hfill
     \begin{subfigure}[b]{0.4\textwidth}
         \centering
         \includegraphics[width=\textwidth]{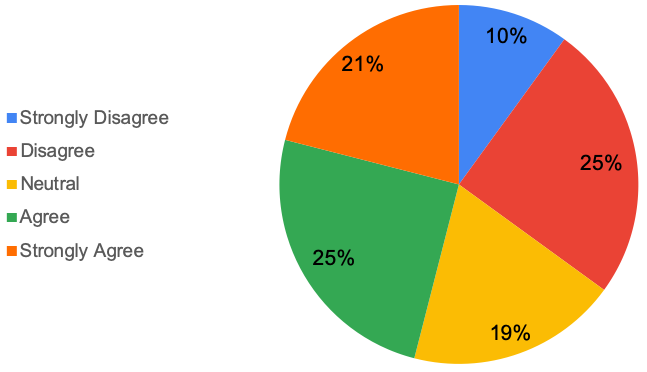}
         \caption{Does the reviewed code handle exceptions?}
         \label{fig:q4}
     \end{subfigure}
     
        \caption{Postmortem questionnaire's results on the code review activity.}
        \label{fig:qs}
\end{figure*}

Figure \ref{fig:qs} shows the results of the questionnaire (full responses can be accessed from \cite{Results:CodeReview}, in \textit{QuestionnaireResponses} folder).
This questionnaire and the results for the code review (table \ref{table:commentscount}) can be used to check how students are performing in the course.
For instance, from table \ref{table:commentscount}, we can observe that around 45\% of the comments for \texttt{Design Decisions}, \texttt{API Design} and \texttt{Architecture} are tagged as \texttt{[CLOSED]}, and from figure \ref{fig:q1}, 61\% agree that the reviewed code is modular.
These two readings suggest that in general, one of the objectives of the course was achieved, Modular Programming.
On the other hand, we can observe from table \ref{table:commentscount} that around 55\% of the comments for \texttt{Exception handling} are tagged as \texttt{[TO BE IMPLEMENTED]}, and from figure \ref{fig:q4}, 35\% disagree that the reviewed code properly handle exceptions.
These two readings suggest that more work needs to be done to achieve one of the objectives of the course, Exception Handling, impacting another objective on High-quality Software. Figure \ref{fig:q2} did not let us make clear conclusions regarding the coverage of logging mechanisms in the course. The even ``agree'' vs. ``disagree'' distribution of the responses (41\% vs. 40\%) suggests two possible explanations: students either have no clear knowledge of how a proper logging mechanism looks like, or they have such knowledge but do not apply it. To cover both of the possibilities, we will need to simultaneously increase the amount of theory in the lectures and introduce more practical exercises to the lab sessions.

\section{Lessons learnt}

\label{sec:lessons}

We report on the lessons learnt after implementing the code review in a classroom mentioned in previous sections. These lessons are supported by feedback provided by students via a questionnaire (that can be found in \cite{Results:CodeReview}). Lessons cover aspects related to the technicalities of the code review process, and some suggestions are in regards to improving the work atmosphere to avoid unnecessary difficulties and achieve productive outcomes.

\subsubsection{Discipline}
Students need explicit guidelines to maintain the positivity of the work environment. Some of the participants (being young) did not know how to take the suggestions as feedback and not as criticisms. They emphasise that developers must try to accommodate reviewers' comments as much as possible. Therefore, the argument of \emph{respect} was brought onto the surface as students stated that respect among team members should be maintained at all times.

\subsubsection{Reference Material}
Students need elaborated reference material. The reference material must include a code review document. The instructions regarding the activity must be precise and supported with examples.
    
\subsubsection{Reviewers' related Guidelines}
The number of reviewers should be between three to five. Less than three reviewers might end up in useless or contradicting comments, e.g., one reviewer says \textit{A}, and the other says \textit{no A}. More than five reviewers will make the review session (the actual meeting) too long, affecting the outcome of the process negatively; \textit{reviewer}s must have clarity that their task is to identify problems and provide suggestions for improvement, not solving it for the developers; quantifiable goals should be set prior the activity, and the number of lines of code should be between 200 and 300. Some students reported that 200 LOC is not enough; however, having more LOC will make the review session (the actual meeting) too long, affecting the outcome of the process negatively.


\subsubsection{Motivation to Work}
To maintain a certain level of motivation (among all participants) during the code review process, the provision of review statistics can be useful. At the end of each review session, the \textit{moderator} can check and report whether all issues were resolved, i.e., are all ``TO\_BE\_IMPLEMENTED'' tags addressed?. They could also report on the achieved goals of the activity. Students recommended instructors' involvement to address the issues that are beyond the control of the teams such as, participants not responding on time.

\subsubsection{Feedback Structure}
It is essential to emphasise the need to have balanced feedback. It is not just about pointing out mistakes in the code, but also acknowledging those good and smart implementations (e.g. from the activity in section \ref{sec:caseStudy}:  ``Smart choice to use implemented DS in MySQL database for data queries").

\subsubsection{Clarity upon Grading Schema of the Code Review}
The code review activity should be part of the course grading schema, clear to students and fairly distribute (the risk of unfair grades' distribution, if not managed, may harm the students' motivation). This will not just play well in students' motivation, but also the code review could be used to determine any correlation between the number of bugs found in a student's code and their final grades.

Code reviews should not be placed close to major milestones of the course, for instance, close to a project delivery or final exam. Some students reported (quoted): \emph{``It would be better to do this activity in the time when we don't have to prepare for our finals so that we could dedicate more time on this activity.''} and {``Don't organize it at the end of the semester before finals''}.

\subsubsection{Activity Technicalities}
Educators should give feedback on the spectrum of goals to ensure that they are doable in the available time frame. Educators should also read and give feedback on students' comments to reduce ambiguity.

\section{Conclusions}
\label{sec:conclusions}
Code reviews are widely used in industry. However, freshly graduated students are not prepared to undertake them. This paper describes a process to perform code reviews in an academic environment that seeks, among other things, prepare students for the future. The process defines two sub-processes to be carried out one by the educator of the course and the other one by students of the course. We applied the process to a first year, large size course in Computer Science. We also presented the findings of the activity and described the lessons learnt after the activity. This paper can be used as a guide to implement code reviews in a classroom (examples of code reviews and material can be found in \cite{Results:CodeReview}).

We now summarise the results of the code review activity against the research questions stated in section \ref{sec:introduction}. The summary relies on the following key sources of information:
\begin{itemize}
    \item[$\star$] Students' evaluations of each other's code (figure \ref{fig:qs}).
    \item[$\star$] Students' responses to a postmortem reflection questionnaire that we handed out to them after the code review activity. We received 78 responses in total, which is slightly more than half of the students (all responses can be found in \cite{Results:CodeReview});
    \item[$\star$] Our gained experience.
\end{itemize}

\paragraph{RQ1}: \emph{What are the benefits and pitfalls for a course performing code reviews?}

When the goals of the code review activity are aligned to the course's goals, the activity can be used to 
\begin{itemize}
    \item[$\bullet$] assess students. Code review activities can be part of the grading criteria of the course. Scores can be calculated by the engagement of the student during the activity, e.g. the number of comments reported, the number of answers given. Or it could be used in a student peer assessment, in which students assess the work of their peers against set assessment criteria (defined by the Educator). When students act as the assessor, they gain an opportunity to better understand assessment criteria potentially increasing their motivation and engagement \cite{peerassessment:2017}. 
    \item[$\bullet$] keep track of the development of the course objectives. Code reviews can be used as a formative assessment if done repeatedly during the course. Educators can track the course objectives and act according to the findings, improving the learning outcomes of the course. For instance, the code review session described previously helped us spot well-developed and underdeveloped topics of the course. The course does not cover exception handling well enough (figure \ref{fig:q4}); it does cover modular programming in depth (figure \ref{fig:q1}); it trains students to write understandable code (figure \ref{fig:q3});
    \item[$\bullet$] track students' learning. Students need to constantly commit their development to a repository. Educators can check their progress and take action if needed. Educators can also check the comments of the code review session. The latter will help Educators track the learning process of the group involved rather than individuals.
\end{itemize}

\noindent
This will boost the quality of teaching.

Code review is an \textit{active learning} activity. It involves students in the learning process by doing things (phase (3) in figure \ref{fig:codeReview}) and thinking about what they are doing (phase (4) in figure \ref{fig:codeReview}). Students use their own efforts to construct their knowledge, guided by Educators. This makes code review an effective tool in making classrooms more inclusive, students more engaged, improving their critical thinking, in general, increasing students' performance in the course \cite{Freeman:2014}. 

The activity does not come for free. Code reviews can fall into different pitfalls. Educators need to know how to mitigate those problems making them benefits of the activity. Code review is a multistep process that requires considerable time and attention during both the preparation and execution phases. Overlooking possible problems during these phases will result in cascading problems during the execution phase. The likelihood increases with large size courses. An example of possible problems is not having \textit{moderator}s well prepared. This will affect the outcome of the activity as students will not be well guided.

Motivating students is a key factor in the code review process. Failing to do so might reflect on students' final grade of the course. Students should have a clear understanding of the activity and Educators need to make them engage in every step.

\paragraph{RQ2}: \emph{What are the benefits and pitfalls for students participating in code reviews?}

Typically, university core courses focus on technical skills, letting out those soft (people) skills. Code reviews help students developing both hard and soft skills. Students reported that the activity helped them learn to read and understand other people's code; take an external look at their code and thus improve it; practise organising their thoughts clearly; practise working under tight deadlines; and how \emph{not} to write code. They have understood that writing code is for the benefit of humans reading the code (machines do not read source code but low-level code). Some of the soft skills exercised were collaborative and teamwork, give and receive appropriate feedback (how to deal with criticism), and how to  resolve conflicts. Code reviews also prepare students to an industrial environment that widely uses reviews as part of their software development process. 

Some students do not know how to deal with criticism, influencing negatively their motivation. Industrial practitioners of code reviews understand the benefits of the activity from a tremendous experience. On the other hand, first-year bachelor students practising it for the first time instinctively take the outcome as a personal criticism. Educators need to be vigilant to this situation and take immediate action to mitigate the problem, not just because it may reflect on the outcome of the activity, but also because it may affect students' personal success. Some responses to the postmortem questionnaire support this conclusion. As a possible remedy,  the participation of higher-grade students, TAs or Educators can be used to make their verdicts in possible conflicts. 

\textit{Owner}s of the code may fall prey to the \textit{reviewer}s' low discipline. Motivation plays a key role once again. A motivated \textit{reviewer} might give a substantial feedback to the \textit{owner}, unlike a unmotivated one. One of the answers we received on this regards reads: \emph{``It would be much better if \textit{reviewer}s have not started two hours before the deadline''}. 

There is a risk of one student damaging the grade of another one during the code review activity. Especially, if \textit{student peer assessment} is exercised. This should be carefully managed as it may impact both students' final grades and motivation.  

\paragraph{RQ3}: \emph{What aspects of the artifacts under review do the students emphasise most?}

They mostly emphasise code-level aspects, such as coding style. This goes in accordance to code reviews practised in industry. Typically, code reviews in industry focus on small changes to the actual software. Educators could, however, take advantage of the activity to let students think and comment about more high level aspects. For instance, code reviews can be exercised as early as the requirement elicitation phase of the software development. Here, the artifact to be reviewed is not a piece of code but a text describing the requirements of the software. This will help students think about the high level structure of the system, let them see a bigger picture and practise reviewing architecture decisions in addition to code.

\bibliographystyle{plain}
\bibliography{bibl}

\end{document}